\documentstyle[12pt]{article}
\begin{document}
\begin{center}
{\large A NOVEL BRST APPROACH IN GENERALIZING THE JACKIW-NAIR ANYON}

\bigskip\bigskip\bigskip\bigskip
Subir Ghosh \\
Physics Department, Gobardanga Hindu College \\
24 Pgs. (North), West Bengal, India\\
\bigskip\bigskip
S. Mukhopadhyay \\
Saha Institute of Nuclear Physics\\
1/AF Bidhannagar, Calcutta 700064, India.

\bigskip\bigskip\bigskip\bigskip
Abstract \\
\end{center}

\noindent A novel BRST quantization is described, which is applied in
generalizing the Jackiw-Nair construction of anyon. We have explicitly
shown that the matter states connected to an unconventional ("non-zero")
BRST ghost sector are physical. They are identified to the Jackiw-Nair system
in a particular gauge. Also for the first time an indepth analysis of the
present kind for a reducible constraint system, ( where the constraints are not
independent), has been performed.

\newpage
In this Letter, we have developed a relativistic Lagrangian model of anyon,
which
is a {\it generalization} of the Jackiw and Nair (JN) construction \cite{JN}.
{\it In a particular gauge our
theory reduces to that of the JN anyon}. We have performed a BRST quantization
of
our \cite{SG} original spinning particle model of anyon, along the lines of
Marnelius
\cite{M1}. In the process we have introduced a novel
idea of utilizing {\it an unconventional BRST ghost sector}. Indeed, this new
scheme is applicable in a much wider class of physical models, although some
formal details have to be worked out. Apart from these two major results, there
are some
more interesting aspects of a rather technical nature : we have carried through
the quantization \cite{M1} in a qualitatively different and also much more
involved
type of system \cite{SG}, which is known as a {\it reducible constrained
system}
\cite{BF}, meaning that the constraints are not independent. This reducibility
crucially affects the final results and has compelled us to use the more
rigorous
prescriptions of \cite{BF} to compute the BRST charge.

What is the motivation behind constructing new models of anyon? The
conventional
model \cite{W} has been that of an interacting theory of a statistical gauge
field
(with Chern-Simmons dynamics) coupled with point particles (in quantum
mechanics)
or with bosonic or fermionic matter fields (in field theory). Elimination of
the
gauge field leaves the matter particle endowed with anyonic spin and
statistics.
 However, the Chern-Simmons construction is non-relativistic and essentially
classical. It has been argued \cite{HKP} that residual effects remain at the
quantum level
even after removal of the gauge field.
All these have motivated the search for minimal relativistic models of anyons
where the
complications due to an interacting theory are absent. Some of the successful
models are \cite{JN,PL,SG}.

Let us briefly recapitulate our model \cite{SG}. It is represented by a point
particle,
with a rotating frame attached to it moving along a world line.  The
instantaneous position and orientation
of the particle with respect to a space fixed frame is denoted by a Poincare
group element ($x^\mu(\tau)$, $\Lambda^{\mu\nu}(\tau)$), where $\tau$ is some
parameter.
Our metric is $g^{00}=-g^{11}=-g^{22}=1$. Linear and angular
velocities are given by $u^\mu ={dx^\mu\over d\tau}$ and $\sigma^{\mu\nu} =
(\Lambda^T {d\Lambda\over d\tau})^{\mu\nu}$. The reparametrization invariant
Lagrangian is
\begin{equation}
L = (M^2 u^2 + {1\over 2}J^2 \sigma^2 + MJ\epsilon^{\mu\nu\lambda}
\sigma_{\mu\nu}
u_\lambda )^{1\over 2},
\end{equation}
where $M$, $J$ are parameters having physical significance of mass and spin
respectively. The canonically conjugate linear and angular momenta
$P_\mu$ and $S_{\mu\nu}$ are related by the constraints which could be
regrouped
in the form
\begin{equation}
\Psi_1 = \epsilon_{\mu\nu\lambda} S^{\mu\nu} P^\lambda -2MJ  \approx 0,~~
\Psi_2 = P^2 - M^2 \approx 0,~~
{\overline \Psi_{(0)\mu}} = S_{\mu\nu} P^\nu \approx 0,
\end{equation}
of which the third one is a second class set. The Marnelius construction is
best
suited for First Class Constraints (FCC), which form a closed algebra. Hence we
modify $\bar\Psi_{(0)\mu}$ \cite{NQ} by
$\Psi_{(0)\mu} = S_{\mu\nu} (P^\nu + \Lambda^{0\nu} )$ and the resulting set of
FCC
$\Psi_A, A=0,1,\mu$, satisfy the commutation relations,
\begin{equation}
[\Psi_{(0)\mu}, \Psi_{(0)\nu}] = iS_{\mu\nu}\Psi_2 - i(\Psi_{(0)\mu}P_\nu -
\Psi_{(0)\nu}P_\mu),
\end{equation}
\begin{equation}
[\Psi_{(0)\mu}, \Psi_1] = 2i\epsilon_{\mu\nu\lambda}P^\nu\Psi^{(0)\lambda},
\end{equation}
A gauge fixing $P_\mu - \Lambda^0_\mu = 0$ gives the original theory back.
This set is not independent (namely reducible)because of the identity
\begin{equation}
\Psi_{(0)\mu} (P^\mu + \Lambda^{0\mu}) = 0.
\end{equation}
This feature of reducibility also agrees with the observation in \cite{JN}.

We now follow the prescription of \cite{M2} for the BRST analysis. The
constraints $\Psi_A$
are all bosonic and we introduce the auxiliary fields:
bosonic Lagrangian multipliers
$(\pi_A, \lambda^A)$ and fermionic ghost antighost pairs,
$(\bar{\cal P}_A, {\cal C}^A)$ and  $(\bar{\cal C}_A, {\cal P}^A)$.
For the single reducibility condition
fermionic multipliers $(\pi', \lambda')$ and the ghost antighost pairs
 and $(\bar{\cal P}', {\cal C}')$ and $(\bar{\cal C}',{\cal P}')$ a further set
of extraghosts $(\pi'', \lambda'')$ and $(\bar{\cal C}'', {\cal P}'')$ are
needed.
A generic canonical pair is written as $(p_A, q^A)$
which satisfy the commutation relation $\{q^A, p_B\} = \delta^A_B$ for
fermionic
and $[q^A, p_B] = i\delta^A_B$ for bosonic degrees of freedom. Construction of
the
the quantum BRST charge, although straightforward, is quite involved (the
system
being reducible). Omitting the details we have
\begin{eqnarray}
Q &=& \Psi_1 {\cal C}^1 + \Psi_2 {\cal C}^2 + \Psi_{(0)\mu} {\cal C}^{(0)\mu}
          -{i\over 2}S_{\mu\nu}{\cal C}^{(0)\mu}{\cal C}^{(0)\nu}\bar{\cal P}_2
\nonumber\\
          &-&2iP_\mu\epsilon_{\mu~\beta}^{~\alpha}\bar{\cal P}_{(0)\alpha}
          {\cal C}_{(0)}^\beta {\cal C}^1 + iP_\mu\bar{\cal P}_{(0)\nu}{\cal
C}_{(0)}^\nu
          {\cal C}_{(0)}^\mu \nonumber \\ &+& (P_\mu - M\Lambda^0_{~\mu}){\cal
C}'\bar{\cal P}'
         {\cal C}_{(0)}^\mu + i(P^\mu + M\Lambda^{0\mu})\bar{\cal
P}_{(0)\mu}{\cal C}'
\nonumber \\
         &+& {1\over2}\bar{\cal P}_2(\bar{\cal P}_{(0)\mu}{\cal C}_{(0)}^\mu -
i{\cal C}'
         \bar{\cal P}').
\end{eqnarray}
This is same as that for the massive spinning particle in (3+1)dimension
\cite{NQ}
except for the first term which is due to the extra constraint. One can check
the
identity $Q^2=0$. The physical state must satisfy $Q|phys> =0$.

The other crucial operator, the ghost number $N_g$ is
\begin{equation}
N_g = {1\over2}({\cal C}^A\bar{\cal P}_A - \bar{\cal P}_A{\cal C}^A)
              -i({\cal C}'\bar{\cal P}' - \bar{\cal P}'{\cal C}')
\end{equation}
The operator ordering ambiguity is removed by following the prescription of
\cite{M1}.
This assigns the ghost number of any operator by $[N_g,{\cal O}] = n_g{\cal
O}$,
where $n_g$ is the ghost number.

Now we start building the Hilbert space, concentrating first on the ghost
sector.
The vacua and the states are lebelled as ${\cal O}|>_{\cal O} = 0$ and $|n_g>$.
The fermionic sector reads
\begin{equation}
|-(+){1\over 2}>_i = |0>_{\bar{\cal P}_i({\cal C}^i}) \sim  \bar{\cal
P}_i({\cal C}^i |0>_{{\cal C}^i(\bar{\cal P}_i)},
{}~~~{\hphantom >}_i<m|n>_j = \delta_{ij}\delta_{m+n}, ~~~ i,j = 0,1
\nonumber \end{equation}
For the other set $(\bar{\cal P}_{(0)\mu}, {\cal C}_{(0)}^\mu )$ we have
\begin{equation}
|-{3\over 2}> = |0>_{{\bar{\cal P}}_{(0)\mu}} ,
|-{1\over 2}>^\mu = {\cal C}_{(0)}^\mu|0>_{\bar{\cal P}_{(0)\mu}},
\end{equation}
\begin{equation}
|+{1\over 2}>_\mu = {1\over2} \epsilon_{\mu\nu\lambda}~
{\cal C}_{(0)}^\nu{\cal C}_{(0)}^\lambda|0>_{\bar{\cal P}_{(0)\mu}},
\end{equation}
\begin{equation}
|+{3\over 2}>_\mu = {1\over6}\epsilon_{\mu\nu\lambda}
{\cal C}_{(0)}^\mu{\cal C}_{(0)}^\nu{\cal C}_{(0)}^\lambda|0>_{\bar{\cal
P}_{(0)\mu}},
\nonumber\end{equation}
\begin{equation}
{\hphantom >}_\nu<m|n>^\mu = \delta^\mu_\nu\delta_{m+n}.
\nonumber\end{equation}
We can interchange ${\cal C}$ and ${\cal P}$ and that will flip the sign of
the ghost number. The states corresponding to other ghosts could be constructed
in this way.

However the bosonic sector consisting of ghosts, (as well as matter states),
is infinite, unlike the fermionic one, since one can operate any number of
bosonic creation
operators on the vacuum. Thus there are a number of possible ghost states,
which
is not correct since there can not be physical ghost excitations \cite{M3}. A
natural way
of removing this degenaracy is to restrict the space to states of {\it finite}
norm
only \cite{M1}. Since inner products such as $<p'|p> = \delta(p'-p)$ are
singular, a
non-hermition state space is forced on us, where bras and kets are not
hermition
conjugate of each other.
For its relations with the conventional Hilbert space
and other details see references \cite{M2}.
Consider the bosonic ghost to start with.
Thus there are two possible `ground states' $|0>_{\bar{\cal P}'}(|0>_{{\cal
C}'})$ which are
annihilated by $\bar{\cal P}'({\cal C}')$ and a pair of towers of states can be
built
by operating ${\cal C}'(\bar{\cal P}')$ repeatedly on them. So the Hilbert
space
consists of the following states.
\begin{equation}
|n> = {({\cal C}')^{n-1}\over \sqrt{(n-1)!}} |1>_{\cal P'} ,~~~
{\hphantom >}_{\cal P'}\! <0|0>_{\cal P'}=1, \nonumber\end{equation}
\begin{equation}
|-n> = {({\cal P}')^{n-1}\over \sqrt{(n-1)!}} |-1>_{\cal C'} ,~~~
{\hphantom >}_{\cal C'}\!<0|0>_{\cal C'}=1.
\end{equation}
The construction of bosonic matter Hilbert space follows in a similar way. One
has
different choices for bra and ket states. The generic ket states are given by
operating an arbitrary
function of the position operator on the momentum `ground state' (i.e. $|0>_P$
with $P_\mu|0>_P = 0$)
given by $|\Phi_{\alpha\beta\cdots}> = \Phi_{\alpha\beta\cdots}(x)|0>_p$ where
the bra states are given by $<x'|C_{\alpha\beta\cdots}$ where
$C_{\alpha\beta\cdots}$
is a covariant constant. For the spin sector one should use any conventional
representation \cite{LIP} and for that sector bra and ket states are hermitian
conjugates of each other.
Such a description of the matter Hilbert space is termed as the wave function
(Schrodinger) representastion \cite{M1,HT}.

Finally we have reached the most interesting part of our work:
finding out the physical matter sector. Firstly we assumed that $|phys>$ is a
direct product of the form $|matter>|ghost>$. Secondly the ghost sector
consists
of a sum of eigenspaces of $N_G$ with real ghost numbers. Hence $|phys>$ is
expanded
in powers of ghosts \cite{HT},
\begin{equation}
|phys> = |M> + |M>_i|G>_i + |M>_{ij}|G>_{ij} + \cdots
\end{equation}
where the number of indices in $|G>$ indicates the number of ghost operators on
vacuum. The states $|M>_{ij\cdots}$ belong to the Dirac Hilbert space \cite{D}.
Normally the condition $Q|phys> = 0$ picks up the $|M>$ connected to zero ghost
sector
$|G>$ (namely $N_g|G> = 0$) in case of Fock representation or the extreme
($\pm$)
ghost number sector (namely $N_g|G> = \pm {m\over2}|G>$, $m$ = number of
constraints) in
the wave function representation, which is relevant here. However the following
comment
in \cite{HT} is worth quoting :"The existence of non-trivial BRST cohomology at
ghost number $\not= \pm {m\over 2}$depends on the detailed structure of the
theory.
It will not concern us here because it does not appear to be related to the
states
of the Dirac method without ghosts." We want to emphasize, (as we will
explicitly show),
that in the present case the $|M>$ connected to a $|G>$ whose ghost number is
{\it not}
$\pm{m\over 2}$ is related to the physical states without ghost and this matter
sector
satisfies a generalized set of JN equations \cite{JN}. In fact this is forced
on us as the
matter part with conventional ghost sector turns out to be trivial, once $Q$ is
operated
on it. The cure for "non-zero" ghost number is also given in \cite{HT}. One
should suitably
include the Lagrange multipliers and their conjugate momenta and extraghosts,
which will
change $Q$ and $N_g$ to
\begin{equation}
\bar Q = Q + \pi_A{\cal P}^A + \pi'{\cal P}' +\pi''{\cal P}'' ,~~~
\bar N_g = N_g - {i\over 2}(\lambda'\pi'+\pi'\lambda')
\end{equation}
In this still enlarged space the ghost number of the physical sector just
obtained
is again zero.
For the construction of the field equations we choose a sub-space given by
\begin{equation}
|phys> = |-{1\over2}>_1|-{1\over2}>_2|-{1\over2}>^\alpha|M_\alpha>,
\end{equation}
and its dual subspace which should involve the ghost states with opposite ghost
number is
\begin{equation}
<phys| = <M_\alpha'|^{~\alpha}\! <{1\over2}|_{~1}\!<{1\over2}|_{~2}\!<{1\over
2}|,
\end{equation}
where $<M|^\alpha$ and $|M'>_\alpha$ are the matter states represented in the
wave
function sector. Note that in $|phys>$ we have retained the relevant variables
only. Imposing the physicality condition $Q|phys> = 0$ and
$<phys|Q = 0$ one gets the following restrictions on the matter space.
\begin{eqnarray}
(\Psi_1\delta_\alpha^\beta - 2i\epsilon^{\lambda ~\beta}_{~\alpha}P_\lambda
|M_\beta> &=& 0, ~~~~
\Psi_2 |M_\mu> = 0, \nonumber\\
\epsilon^{\mu\nu\lambda}(\Psi_\nu + iP_\nu)|M_\lambda> &=& 0,~~~~
(P_\mu + \Lambda^0_{~~~\mu})|M^\mu> = 0, \label{c}\\
<{M'}^\mu| (\Psi_\mu + 2iP_\mu) &=& 0.
\end{eqnarray}
It is rather inconvenient to have both the bra states and the ket states
restricted.
We would like to have the bra states to be unrestricted. So we write the
bra state as a projection on an unrestricted state by the projection operator
$\Pi_\alpha^\beta = (\delta_\alpha^\beta - V_\alpha V^{-2} V^\beta)$ where
$V_\alpha = \Psi_\alpha
+ 2iP_\alpha$ and we have written $V^{-2}$ formally. Since to get the field
equations only the inner products are relevant one can replace $<{M'}_\alpha|$
and $|M_\alpha>$ by $<\tilde M_\alpha|$ , $|\tilde M_\alpha>$ where $<M_\alpha|
=
<{\tilde M}^\beta|\Pi^\beta_{~\alpha}$ and $|{\tilde M}_\alpha> =
\Pi_\alpha^{~\beta}|M_\beta>$. So the bra state remains unrestricted but the
ket state has to satisfy the additional requirement
\begin{equation}
(\Psi_\mu + 2iP_\mu)|M^\mu> =0.\label{d}
\end{equation}
We choose the matter sector states as $|M_\alpha> = \Sigma_n f_{\alpha,n}
(x)|0>_p|n>$ for ket state and $<x'|<n'| = <{M'}_\alpha|$ where $|n>$ is the
representation of the spin sector. Taking the inner product of the equations
(\ref{c},\ref{d}) with these ket states we have
the following equations
\begin{eqnarray}
\Psi_2 f_{\mu,n}(x) &=& 0, ~~~~
((\Psi_1)_{nn'} \delta_\alpha^\beta - 2iP_\mu\epsilon_{~\alpha}^{\mu~\beta}
\delta_{nn'})f_{\beta,n'}(x) = 0, \nonumber\\
((\Psi_\mu)_{nn'} + 2iP_\mu \delta_{nn'})f^\mu_{n'}(x) &=& 0,~~~
\epsilon^{\mu\nu\lambda}((\Psi_\nu)_{nn'} + iP_\nu\delta_{nn'}) f_{\lambda n'}
 = 0, \nonumber\\
(P_\mu\delta_{nn'} + M(\Lambda^0_\mu)_{nn'}) F^\mu_{~~,n'}(x) &=& 0.
\label{n} \end{eqnarray}
Here $P_\mu$is written as an operator $-i\partial_\mu$, $J_\mu$ ($S_{\mu\nu} =
 \epsilon_{\mu\nu\lambda} J^\lambda$)and $\Lambda^0_\mu$ are written as
their representations $<n'|J_\mu|n>$ and $<n'|\Lambda^0_\mu|n>$. This is the
generalized version of the field equation. The matrices representing $J_\mu$
serves the role of
the `gamma matrices' of the anyonic equation.
This generalized set of equations (\ref{n})
constitutes one of the main results of our work. Representation of the
full $J_\mu$, $\Lambda^0_\mu$ system for the gauge symmetric version is
still open. The problem of constructing a
 local field theoretic Lagrangian, the variation of which will reproduce the
above set of equations is under study.

As promised earlier, we show how one can get the JN construction from our
generalized version.
As mentioned earlier we should take a gauge fixing.
To get the JN construction from our equations
the covariant gauge given by $\chi_\mu = P_\mu -
M\Lambda^0_\mu \approx 0$ is prescribed.
A generic gauge fixing requires introduction of extra
terms in $Q$ which might affect the equations of motion. However this
particular
gauge is judicious
in the sense that all the extra terms
involve the auxiliary fields only and so do not modify the set of equations.
After the gauge has been fixed we have to find a suitable representation in
keeping
with the gauge fixing. It should correspond to the arbitrary value of
the spin. A suitable choice is given by the following \cite{JN,LIP}.
\begin{eqnarray}
J^0|\lambda, n> &=& (\lambda + n) |\lambda, n>, \\
J^+|\lambda, n> &=& \sqrt{(2\lambda + n)(n+1)}|\lambda, n+1>, \\
J^-|\lambda, n> &=& \sqrt{(2\lambda + n - 1)n}|\lambda, n-1>,
\end{eqnarray}
with $J^{\pm} = J^1 \mp iJ^2$, $J^-|\lambda, 0> = 0$, $J^2|\lambda, n> =
\lambda(\lambda+1)|\lambda, n>$.
This is the representation bounded from below and one can get the
representation
bounded from above by replacing $J$ by $-{\tilde J}$. Another point to
note is that all the operators that can be obtained by changing the operator
orderings
in $\Psi_{(0)\mu}$ are reduced to the same expression under the covariant
gauge fixing. So there is an ambiguity in the equation in the gauge fixed
form. To resolve the ambiguity we are following the prescription that we
first write down the equation in terms of the symmetrically ordered form
\begin{equation}
\Psi_\mu^{(s)} = {1\over2}[S_{\mu\nu}(P^\nu + M\Lambda^{0\nu}) +(P^\nu +
M\Lambda^{0\nu})S_{\mu\nu}],
\end{equation}
and then we fix the gauge.
Now substitution of this representation will
lead to the JN construction. We emphasize that in our analysis the mass shell
condition, the Pauli Lubanski scalar equation, the subsidiary condition as well
as the transversality condition of \cite{JN} {\it all} appear in an equal
footing ( that is the latter
are not invoked from the outside ).

Hence we conclude by noting that we have derived a set of generalized anyon
wave equation which in a
particular gauge is identical to the full JN construction. A novel BRST
analysis
which has led to the above construction explicitly
demonstrates that the matter states connected to unconventional ghost sector
can indeed be related to the Dirac matter sector without ghosts. In the process
the Marnelius construction has been carried through in a reducible constraint
system.

One of us (S. M.) would like to acknowledge the financial support of Council
of Scientific and Industrial Research, India.

\end{document}